\documentclass{article}
\usepackage{graphicx,cite}
\catcode`\@=11

\@addtoreset{equation}{section}
\def\theequation{\arabic{section}.\arabic{equation}}
     
\catcode`\@=11
\def\thesection{\arabic{section}.}

\def\appendix{\setcounter{section}{0}
        \def\thesection{Appendix.}
        \def\theequation{\Alph{section}.\arabic{equation}}}
\def\section{\@startsection{section}{1}{\z@}{3.5ex plus 1ex minus
   .2ex}{2.3ex plus .2ex}{\large\bf}}
     
\def\eqnarray{\let\@currentlabel=\theequation\refstepcounter{equation}
    \global\@eqnswtrue
    \global\@eqcnt\z@\tabskip\@centering\let\\=\@eqncr
    $$\halign to \displaywidth\bgroup\@eqnsel\hskip\@centering
      $\displaystyle\tabskip\z@{##}$&\global\@eqcnt\@ne 
       \hfil${{}##{}}$\hfil
      &\global\@eqcnt\tw@ $\displaystyle\tabskip\z@{##}$\hfil 
       \tabskip\@centering&\llap{##}\tabskip\z@\cr}
\def\lefteqn#1{\hbox to 4\arraycolsep{$\displaystyle #1$\hss}}
\long\def\@makefntext#1{\parindent 0cm\noindent
\hbox to 1em{\hss$^{\@thefnmark}$}#1}

\newcommand{\captionfonts}{\small}
\makeatletter  
\long\def\@makecaption#1#2{%
  \vskip\abovecaptionskip
  \sbox\@tempboxa{{\captionfonts #1: #2}}%
  \ifdim \wd\@tempboxa >\hsize
    {\captionfonts #1: #2\par}
  \else
    \hbox to\hsize{\hfil\box\@tempboxa\hfil}%
  \fi
  \vskip\belowcaptionskip}
\makeatother   

\begin{document}
\begin{titlepage}
\vspace{.5in}
\begin{flushright}
gr-qc/0702094\\
January 2007\\
\end{flushright}
\vspace{.5in}
\begin{center}
{\Large\bf
 Black Hole Entropy and the Problem of Universality}\\
\vspace{.4in}
{S.~C{\sc arlip}\footnote{\it email: carlip@physics.ucdavis.edu}\\
       {\small\it Department of Physics}\\
       {\small\it University of California}\\
       {\small\it Davis, CA 95616}\\{\small\it USA}}
\end{center}

\vspace{.5in}
\begin{center}
{\large\bf Abstract}
\end{center}
\begin{center}
\begin{minipage}{4.75in}
{\small
A key test of any quantum theory of gravity is its ability to reproduce 
the known thermodynamic properties of black holes.  A statistical
mechanical description of the Bekenstein-Hawking entropy once seemed 
remote, but today we suffer an embarrassment of riches: many different 
approaches to quantum gravity yield the same entropy, despite counting 
very different states.  This ``universality'' suggests that some 
underlying feature of the classical theory may control the quantum 
density of states.  I discuss the possibility that this feature is an 
approximate two-dimensional conformal symmetry near the horizon.
}
\end{minipage}
\end{center}
\end{titlepage}
\addtocounter{footnote}{-1} 

\section{Introduction}

In the continuing search for a quantum theory of gravity, black hole 
thermodynamics may be the nearest thing we have to an ``experimental''
result.  While Hawking radiation has not yet been directly observed, 
the derivations of the Hawking temperature \cite{Hawking} and the 
Bekenstein-Hawking entropy \cite{Bekenstein} are robust enough that 
any purported quantum theory that failed to reproduce these results 
would be viewed with deep skepticism.

The Bekenstein-Hawking entropy depends on both Newton's constant $G$
and Planck's constant $\hbar$, and an understanding of the microscopic 
states responsible for this entropy might provide fundamental insights 
into quantum gravity.  A decade ago, the question of what these microstates 
were had a simple answer: ``We don't know.''  Several interesting 
ideas had been proposed---entanglement entropy, for example \cite{Sorkin}, 
or the entropy of quantum fields near the horizon \cite{tHooft}---but 
none seemed adequate.  

Today, by contrast, we suffer from an embarrassment of riches.  The 
Bekenstein-Hawking entropy of a black hole can apparently be explained 
by
\begin{itemize} 
\item weakly coupled string and D-brane states \cite{StromVafa,Peet};
\item horizonless ``fuzzball'' geometries \cite{Mathur};
\item states in a dual conformal field theory ``at infinity'' \cite{AGMOO};
\item spin network states at the horizon \cite{Ashtekar};
\item spin networks \emph{inside} the horizon \cite{Livine};
\item ``heavy'' degrees of freedom in induced gravity \cite{Fursaev};
\item elements of a causal set in the domain of dependence of a horizon
 \cite{Rideout};
\item inherently global characteristics of a black hole spacetime 
 \cite{Hawkingb};
\item no microscopic states: the entropy can be obtained from quantum 
 field theory in a fixed background, with no information about quantum 
 gravity needed \cite{Hawking}.
\end{itemize}
None of these explanations is complete---string theory, for example,
typically gives exact answers only for supersymmetric black holes
and their duals, while loop quantum gravity requires a choice of a
poorly understood adjustable parameter.  But within their domains of 
applicability, all seem to work.  This is a puzzle: even if one rejects 
the existing approaches to quantum gravity, one must still explain why 
Hawking's original calculation, which required no detailed assumptions 
about quantum gravity, should agree with the enumeration of states in
\emph{any} quantum theory. 

To put this puzzle in context, it is useful to compare ``ordinary''
thermodynamics.  The quantum theory of an ideal gas allows us to specify 
and count states, and the resulting entropy agrees, up to typically
small corrections, with the classical prediction.  But this is to be
expected: the correspondence principle relates the quantum states to
the classical phase space, forcing an approximate agreement between 
the two theories.  A classical black hole, on the other hand, has 
no hair---there \emph{is} no classical phase space to explain the 
thermodynamics.  The states responsible for black hole entropy must 
be fundamentally quantum mechanical, and there is no obvious reason 
for them to have any preconceived behavior.

The ideal gas analogy suggests, however, that perhaps some \emph{different}
classical property of general relativity might control the density of
quantum states.  It is thus natural to search for some classical aspect 
of general relativity---perhaps a symmetry---that might govern the 
state-counting of quantum gravity.

\section{Conformal symmetry and the Cardy formula}

There is, to the best of my knowledge, only one known case in which a 
classical symmetry determines a universal form for the density of states 
of a quantum theory.  A two-dimensional conformal field theory is 
characterized by two symmetry generators $L[\xi]$ and ${\bar L}[{\bar\xi}]$, 
which generate holomorphic and antiholomorphic diffeomorphisms.  The 
Poisson bracket algebra of these generators is given by the unique 
central extension of the group of two-dimensional diffeomorphisms, the 
Virasoro algebra:
\begin{eqnarray}
\left\{L[\xi],L[\eta]\right\} &=& L[\eta\xi' - \xi\eta']
  + \frac{c}{48\pi}\int dz
  \left( \eta'\xi^{\prime\prime} - \xi'\eta^{\prime\prime}\right)\nonumber \\
\left\{{\bar L}[{\bar\xi}],{\bar L}[{\bar\eta}]\right\} 
  &=& {\bar L}[{\bar\eta}{\bar\xi}' - {\bar\xi}{\bar\eta}']
  + \frac{{\bar c}}{48\pi}\int d{\bar z}
  \left( {\bar\eta}'{\bar\xi}^{\prime\prime} 
  - {\bar\xi}'{\bar\eta}^{\prime\prime}\right) \label{a1}\\
\left\{L[\xi],{\bar L}[{\bar\eta}]\right\} &=& 0 \ ,\nonumber
\end{eqnarray}
where the central charges $c$ and $\bar c$ (the ``conformal anomalies'') 
depend on the particular theory.  The zero-mode generators $L_0 =
L[\xi_0]$ and ${\bar L}_0 = {\bar L}[{\bar\xi}_0]$ are conserved
charges, roughly analogous to energies.  

In 1986, Cardy discovered a remarkable characteristic of such theories 
\cite{Cardy,Cardyb}.  Given \emph{any} two-dimensional conformal 
field theory for which the lowest eigenvalues $\Delta_0$ of $L_0$ and 
${\bar\Delta}_0$ of ${\bar L}_0$ are nonnegative, the asymptotic density 
of states at large eigenvalues $\Delta$ and $\bar\Delta$ takes the
form
\begin{equation}
\ln\rho(\Delta,{\bar\Delta}) \sim 
  2\pi\sqrt{\frac{(c-24\Delta_0)\Delta}{6}} 
  + 2\pi\sqrt{\frac{({\bar c}-24{\bar\Delta}_0){\bar\Delta}}{6}} \ .
\label{a2}
\end{equation}
Higher order corrections to this relation are also uniquely determined
\cite{Carlipl,Farey,Birminghama}.  The entropy is thus fixed by 
symmetry, independent of any details of the states being counted.

Black holes are not, of course, two-dimensional, and neither are they
conformally invariant.  But there is a sense in which the near-horizon
region of a black hole is \emph{almost} two-dimensional and \emph{almost}
conformally invariant.  For example, a quantum fields near the horizon 
can be approximately described by two-dimensional conformal field theory
\cite{Birmingham,Gupta,Camblong}; indeed, such a description has recently
been shown to determine the Hawking temperature, by a procedure closely 
related to the computation of the conformal anomaly \cite{Wilczek,Iso}.  
Further, the surface gravity, and therefore the Hawking temperature, 
of a black hole may be expressed in a conformally invariant manner 
\cite{Jacobson}, and a generic black hole metric admits an approximate 
conformal Killing vector near the horizon \cite{Martin,Martinb}.  So 
it is at least possible that results from two-dimensional conformal 
field theory may be relevant.

\section{The BTZ black hole}

The first demonstration that conformal field theory techniques could 
explain black hole thermodynamics came from the study of the BTZ black 
hole \cite{BTZ,BHTZ,Carlipa,Carlipb}, a three-dimensional, asymptotically
anti-de Sitter black hole with cosmological constant $\Lambda=-1/\ell^2$
and metric
\begin{eqnarray}
ds^2 &=& -( N^\perp)^2dt^2 + f^{-2}dr^2
  + r^2\left( d\phi + N^\phi dt\right)^2 \label{a3}\\
&&\hbox{\small with}\ \ N^\perp = f
  = \left( -8GM + \frac{r^2}{\ell^2} + \frac{16G^2J^2}{4r^2} \right)^{1/2},
  \quad N^\phi = - \frac{4GJ}{r^2} \ .\nonumber
\end{eqnarray}
Although the spacetime described by (\ref{a3}) is one of constant negative
curvature, it is a genuine black hole, with a Carter-Penrose diagram 
virtually identical to that of the usual four-dimensional Schwarzschild-AdS 
black hole.  The BTZ black hole has an event horizon at $r_+$ and an inner
Cauchy horizon at $r_-$, with
\begin{equation}
M = \frac{r_{+}^{2} + r_{-}^{2}}{8G\ell^{2}} \ , \qquad
J = \frac{r_{+}r_{-}}{4G\ell} \ .\label{a3a}
\end{equation} 
Most important for our purposes, the BTZ black hole exhibits conventional 
thermodynamic behavior, with an entropy 
\begin{equation}
S = \frac{2\pi r_+}{4\hbar G}
\label{a4}
\end{equation}
equal to a quarter of its horizon size.

The existence of such thermodynamic properties is, at first sight, a 
mystery: general relativity in three spacetime dimensions has no local 
degrees of freedom \cite{Carlipc}, and there seems to be no room for 
microscopic states to account for the entropy.  A partial resolution 
was discovered independently by Strominger \cite{Strominger} and 
Birmingham, Sachs, and Sen \cite{BSS}.  The asymptotic boundary of 
three-dimensional anti-de Sitter space is a flat two-dimensional cylinder, 
so it is not too surprising that the diffeomorphisms that preserve the 
asymptotic behavior of the BTZ metric form a Virasoro algebra.  It is, 
perhaps, surprising that this algebra has a classical central charge, 
but Brown and Henneaux showed in 1986 that it does \cite{BH}, with
\begin{equation}
 c = {\bar c} = \frac{3\ell}{2G}\quad \hbox{and} \quad
\Delta = \frac{1}{16G\ell} (r_+ + r_-)^2 \ , \quad 
{\bar\Delta} = \frac{1}{16G\ell} (r_+ - r_-)^2 \ .
\label{a4a}
\end{equation}
Substituting these values into the Cardy formula (\ref{a2}), and
assuming that $\Delta_0={\bar\Delta}_0=0$, one easily obtains the 
entropy (\ref{a4}).

The entropy thus seems to be related to ``boundary conditions'' at 
infinity.  We can gain further insight by considering the Chern-Simons
formulation of three-dimensional gravity \cite{Achucarro,Witten}.  While
a Chern-Simons theory is gauge-invariant on a \emph{compact} manifold,
the presence of a boundary breaks this symmetry.  The result is a
Goldstone-like mechanism, in which ``would-be pure gauge'' degrees
of freedom become physical at the boundary \cite{Carlipe}, leading
to a dynamical Wess-Zumino-Witten model, in this case at infinity
\cite{EMSS,Carlipm}.  A similar emergence of Goldstone-like dynamical 
degrees of freedom can be demonstrated, with a bit more difficulty,
in the metric formalism \cite{Manvelyan,Carlipd}.  Whether these
degrees of freedom can actually explain the entropy (\ref{a4}) remains 
an open question---for a review, see \cite{Carlipb}---but they are 
certainly good candidates.

\section{How to ask the right question}

While these results for the BTZ black hole are certainly suggestive,
they are also clearly not good enough.  The Chern-Simons formulation
of gravity is possible only in three dimensions, and it is only in
three dimensions that the boundary will admit a nice two-dimensional 
description.  While many string theoretical black holes have a near-horizon 
structure that looks like that of a BTZ black hole \cite{Skenderis}, 
allowing the use of the results of the preceding section, most black 
holes do not have such nice characteristics.

Moreover, it would be useful to find black hole microstates at the 
horizon rather than at infinity.  Although a ``holographic'' description 
at infinity may be possible, such a description would make it rather 
difficult to disentangle degrees of freedom in a spacetime with more 
than one black hole.  There are, for example, multi-black hole solutions 
in three dimensions with the same asymptotic symmetries as the BTZ black 
hole \cite{Mansson}; a complete microscopic description of black hole 
entropy should surely allow us to distinguish these.

We might thus take away the following lessons from the BTZ black hole: 
\begin{itemize}
\item we should look for ``broken gauge invariance'' to provide
 new degrees of freedom;
\item we should at least \emph{hope} for an effective two-dimensional 
 picture, which would allow us to use the Cardy formula;
\item but we should look near the horizon for our new Goldstone-like 
 modes.
\end{itemize}

To proceed further, we must first address a general but somewhat delicate 
issue: how does one ask a question about a black hole in a quantum theory 
of gravity?  This question is seldom asked, because the ``usual'' answer 
seems obvious: we fix a black hole background, and then ask about quantum 
fields, gravitational perturbations, and the like in that background.  In
a full quantum theory of gravity, however, we cannot do this: there is 
no fixed background, and the uncertainty relations prevent us from simply 
imposing a black hole metric.

A question about a black hole in quantum gravity is a question about a
conditional probability: ``If [some defining characteristic] is present,
what is the probability of [some black hole property]?''  To answer, we
must first choose a defining characteristic (\emph{not} the full classical
metric!), and then figure out how to require its presence.  For the BTZ
black hole computation of the preceding section, for example, the defining 
characteristic was the asymptotic behavior of the metric.  We may instead 
impose boundary conditions at past null infinity ${\cal I}^-$ \cite{Cadoni}. 
To search for horizon degrees of freedom, we might impose conditions on 
the near-horizon behavior of the metric at constant time (see, for example, 
\cite{Carliph,Carlipi,Navarro,Cvitanb}), or on the behavior of the metric 
near a null ``isolated horizon'' (see, for example, \cite{Solo,Carlipj,%
Giacomini,Carlipf}).  In each case, the hope is that such conditions
break gauge (or diffeomorphism) invariance in a manner that leads to
new degrees of freedom.

I know of two ways to impose such conditions.  One, first introduced 
in \cite{Carlipe}, is to treat the horizon as a sort of ``boundary'' 
at which suitable boundary conditions are prescribed.  In a path integral 
approach, for instance, we can split spacetime into two regions along 
a hypersurface $\mathcal H$ and perform separate path integrals over 
fields on each side, with fields at $\mathcal H$ restricted by our 
horizon conditions.  Such split path integral has been studied in detail 
in three dimensions \cite{Wittenc}, where it leads to the same WZW models 
that were discussed in the preceding section.  The horizon is not, of 
course, a genuine boundary, but it is a hypersurface upon which we impose 
``boundary conditions,'' and this turns out to be good enough.

For the black hole, this ``horizon as boundary'' approach has been shown 
to correctly reproduce the Bekenstein-Hawking entropy by way of the Cardy 
formula \cite{Carliph,Carlipi,Navarro,Cvitanb}.  On the other hand, the 
diffeomorphisms whose algebra yields that central charge---essentially 
those that leave the lapse function invariant---are generated by vector 
fields that blow up at the horizon, and the significance of this divergence 
is not yet clear \cite{Dreyer,Koga,Pinamonti,Huang,Bergamin,Kogab,Kogac}.  
It has also been difficult to apply these methods directly to the 
two-dimensional black hole.

A second, newer, approach is to literally impose the desired ``defining 
characteristics'' of the black hole as added constraints in the canonical 
theory.  We can then use the standard tools of constrained Hamiltonian 
dynamics \cite{Dirac,Diracb,Bergmann} to study the resulting model.  
This approach was first proposed in \cite{Carlipf}, and has had some 
successes, but is still in its early stages.  In the following section, 
I will briefly illustrate it for the two-dimensional dilaton black hole.

\section{Dilaton gravity with horizon constraints}

Two-dimensional dilaton gravity is described by the action
\begin{equation}
I = \frac{1}{16\pi G}\int d^2x\sqrt{g}\left(\varphi R + V[\varphi]\right)
\label{a5}
\end{equation}
with a metric $g_{ab}$ and a scalar dilaton $\varphi$ with an arbitrary
potential $V[\varphi]$ \cite{Kunstatter,Grumiller}.  This action appears 
in many contexts; in particular, it can be obtained by dimensionally 
reducing higher-dimensional general relativity, in which case $\varphi$ 
has an interpretation as the ``transverse area.''  Although the restriction
to two dimensions is a strong one, the action (\ref{a5}) arguably describes 
the generic behavior of a black hole near enough to the horizon.

Let us now look at ``Euclidean'' dilaton gravity, and consider the method
of ``radial quantization'' used in string theory.  That is, we take the 
metric to be positive definite,
\begin{equation}
ds^2 = N^2f^2 dr^2 + f^2\left(dt + \alpha dr\right)^2 \ ,
\label{a6}
\end{equation}
and evolve radially outward from the origin.  Using standard canonical
methods, it is straightforward to check that the momenta conjugate to 
$f$ and $\varphi$ are
\begin{equation}
\pi_f = \frac{1}{Nf}\left(\varphi' - \alpha{\dot\varphi}\right) \ ,\qquad
\pi_{\varphi} = \frac{1}{Nf}\left(f' - (\alpha f)^\cdot\,\right) \ ,
\end{equation}
and that the generators of diffeomorphisms are
\begin{equation}
{\cal H}_\perp = f\pi_f\pi_\varphi 
  + f\left(\frac{\dot\varphi}{f}\right)^\cdot - f^2V \ ,\qquad
{\cal H}_\parallel = \pi_\varphi\dot\varphi -f{\dot\pi}_f \ . 
\label{a7} 
\end{equation}
It is also not hard to show that these symmetries form a Virasoro algebra 
(with $c={\bar c}=0$), with
\begin{equation}
L = \frac{1}{2}\left({\cal H}_\parallel + i{\cal H}_\perp\right) \ , \qquad
{\bar L} = \frac{1}{2}\left({\cal H}_\parallel - i{\cal H}_\perp\right) \ .
\label{a8} 
\end{equation}

Let us now impose ``horizon constraints'' on our initial surface.
Teitelboim \cite{Teitelboim} has shown that the condition for the origin 
to be a horizon in Euclidean gravity is that $\varphi' = \dot\varphi = 0$,
or, equivalently, $s = {\bar s} = 0$, where
\begin{equation}
\vartheta = \left(\dot\varphi - if\pi_f\right)/\varphi  
\label{a9}
\end{equation}
is the Euclidean version of the expansion and $s=\varphi\vartheta$.  
This is not quite suitable for our purposes, though, since it gives 
a constraint at a single point rather than an initial surface.  To 
define radial evolution, we must instead start at a ``stretched horizon,'' 
a small circle around the horizon.  The appropriate constraint---determined, 
for example, by the requirement that the Hamiltonian be well-defined 
\cite{Carlipj}---can be written in terms of $s$ and the surface gravity 
$\kappa$ as
\cite{Carlipf,Carlipk}
\begin{equation}
K = s - a\varphi_+(\kappa-\kappa_+) = 0, \qquad
{\bar K} = {\bar s} - a\varphi_+({\bar\kappa}-{\bar\kappa_+}) = 0
\label{a10}
\end{equation}
where
\begin{equation}
\kappa = \frac{\dot f}{f} - i\pi_\varphi ,
\label{a11}
\end{equation}
$\varphi_+$ is the value of the dilaton at the horizon, and $a$ and
$\kappa_+$ are constants.

The constraints (\ref{a10}) are not preserved by the generators $L$
and $\bar L$ of the conformal symmetry---that is, Poisson brackets
such as $\{L[\xi],K\}$ do not vanish.  We can fix this, though, by
using a trick first introduced by Bergmann and Komar to handle second
class constraints \cite{Bergmann}: we add ``zero,'' in the form of
multiples of $K$ and $\bar K$, to the symmetry generators to make
their brackets with the constraints vanish.  A fairly straightforward
computation shows that as one approaches the horizon, the modified 
generators again obey a Virasoro algebra, but now with
\begin{equation}
c={\bar c} = 6\pi a\varphi_+ \ , \qquad 
\Delta={\bar\Delta} = \frac{\pi a\varphi_+}{4}
   \left(\frac{\kappa_+\beta}{2\pi}\right)^2 \ ,
\label{a12}
\end{equation}
where we assume that our diffeomorphisms are periodic in $t$ with period 
$\beta$.  Smoothness of the metric (\ref{a6}) at the origin requires a
periodicity $\beta=2\pi/\kappa_+$, yielding the usual Hawking temperature.
The Cardy formula (\ref{a2}) then gives an entropy
\begin{equation}
S = \pi a\frac{\varphi_+}{4G}
\label{a13}
\end{equation}

It remains for us to determine the constant $a$.  For a dimensionally
reduced spherically symmetric black hole, 
\begin{equation}
f^2 = 1 - \left(\frac{r_+}{r}\right)^{D-3}, \quad 
\varphi = \varphi_+\left(\frac{r}{r_+}\right)^{D-2} \ ,
\label{a14}
\end{equation}
and an easy computation gives $a=2$. More generally, this value appears
to be determined by demanding that the stretched horizon be affinely 
parametrized, but this issue is not quite settled.  

Given a value $a=2$, the entropy (\ref{a13}) is precisely $2\pi$ times 
the standard Bekenstein-Hawking entropy.  A similar factor of $2\pi$ was 
found in \cite{Park}.  I believe it arises because in radial quantization 
we are computing the entropy over ``all times,'' on an initial surface 
consisting of a circle of circumference $2\pi$.

Up to this factor of $2\pi$, the central charge (\ref{a12}) agrees with
that of the ``horizon as boundary'' approach \cite{Carlipi}, and is
closely related to the BTZ central charge (\ref{a4a}).  At first sight, 
the conformal charges $\Delta$ and $\bar\Delta$ in (\ref{a12}) look 
rather different from the corresponding BTZ quantities in (\ref{a4a}). 
But this difference is easily explained: the asymptotic diffeomorphisms
that generate $\Delta$ and $\bar\Delta$ for the BTZ black hole depend
on $t\pm\ell\phi$, and have periodicities \cite{CarTeit}
\begin{equation}
\beta_\pm = \left(\frac{r_+\pm r_-}{r_+}\right)\beta \ .
\end{equation}
Substituting these expressions into (\ref{a12}), we find differing 
``left'' and ``right'' conformal charges of the same form as (\ref{a4a}).

While these results are intriguing, we still have a long way to go to
establish (or reject) the proposal that black hole thermodynamics can
be understood in terms of horizon symmetries.  Perhaps the most important
open question is whether the ``Goldstone-like'' modes can couple to matter
to produce Hawking radiation.  In three dimensions, this has been shown
to occur \cite{Emparan}: a classical matter source coupled to the
boundary degrees of freedom of the BTZ black hole causes transitions
among boundary states, and detailed balance arguments can then be used
to derive the Hawking radiation spectrum.  For more general black holes,
the anomaly-based computations of Hawking radiation \cite{Wilczek,Iso} 
are suggestive (see also \cite{Parikh,Mattingly}), but the full story
will require more work.

\vspace{1.5ex}
\begin{flushleft}
\large\bf Acknowledgments
\end{flushleft}
This work was supported in part by the U.S.\ Department of Energy 
under grant DE-FG02-99ER40674

\end{document}